# Chaos synchronization between the Ikeda systems coupled in a star network topology

E.M. Shahverdiev

Azerbaijan National Academy of Sciences Institute of Physics

H.Javid Avenue,31 Baku, AZ1143

e-mail: e.shahverdiev@physics.science.az

Abstract

The case of a star topology is studied on the example of Ikeda model –famous interdisciplinary system. The existence and stability conditions for the complete synchronization between constituent systems are found. An agreement between numerical simulations and theory is excellent. Some security implications for the computer networks are underlined.

## I. Introduction

Chaos synchronization is one of the powerful control methods in nonlinear sciences and wide spread in the scientific, technological fields and natural world, etc., see e.g. [1-2] and references there-in.

Synchronization between systems can help to achieve higher power lasers. This is especially important for the synchronization between Terahertz sources. In recent years it was established that synchronization between Terahertz sources could be potentially helpful to achieve mW powers, see e.g. [3] and references there-in. Such powers can be vital in creating adequate powers for practical applications. Synchronization between thousands and thousands Josephson junctions present in high temperature superconductors could be helpful in achieving this goal [3]. Additionally, such synchronization is of immense use to create mobile, small size, cost effective Terahertz devices. Such devises can be used in remote security screening, in detecting fake painting, plastic land mines, etc., see [4] and references there-in.

For the coupled chaotic systems, many different synchronization states have been studied. Complete or identical synchronization [5-6] was the first to be discovered and is the simplest form of synchronization in chaotic systems. Other types of synchronization include: phase synchronization [7]; lag synchronization [7]; inverse synchronization [8-9]; generalized synchronization [10]; cascaded and adaptive synchronization [11]; anticipating synchronization [12-13], etc.

Synchronization in complex systems is of a certain importance in governing and performance improving point of view [14]. This is based on the fact that a chaotic attractor consists of infinite number of periodic orbits, along which the nonlinear system's performance differs. Choosing the "right" periodic orbit the system's yield can be optimized.

As synchronization is associated with communication, a study of existence and stability conditions for synchronization is of paramount importance in networks [1].

While focusing on the positive side of the chaos synchronization, one should not forget about the situations when synchronization between interacting systems could quite harmful. For example, in an epileptic patient synchronization between neurons could be the reason for epilepsy seizures [15]. Anticipating synchronization could be quite helpful for diagnostic purposes, e.g. by anticipating epileptic seizures [12, 13].

Due to the finite speed of information propagation between the interacting systems, feedback and memory effects, etc. time delay systems [16] are wide spread in the science, technology, and in the natural world [1]. Apart from this, these systems can be used to model space-time processes, see e.g. [16] and references there-in. Because time delay systems are in fact infinite dimensional (as initial conditions are given on the interval and number of points are infinite) these systems can be used to describe partial differential equations. Most importantly, time delay systems (functional differential equations) are capable to generate hyper chaos–from the security point of view very attractive property for chaos-based communication systems [17-18].

This paper studies chaos synchronization in a star network topology based on the Ikeda system-paradigmatic model of chaotic dynamics in time delay systems [19-20].

The originality of this paper is in the building a bridge between chaos synchronization and computer network(s). Synchronization is important in chaos-based communication [1]: At the transmitter a message is masked with chaos, then this combined signal is transmitted to the receiver system. At the receiver due to the chaos synchronization between the transmitter and the receiver chaos is regenerated. Deducting the receiver output from the receiver input one can decode the transmitted message. An additional layer of security could be provided by such an approach to communicating data packets between the laser systems and the computers.

In this paper we consider the star topology connection between the Ikeda systems. Both the existence and stability conditions for the complete synchronization are derived. We also provide the results of numerical simulations to support the theory. These findings are of certain importance in communication between computers, in obtaining high power lasers.

The organization of the rest of this paper is as follows. In Section 2 we briefly introduce the working model and present the results of the analytical study. Section 3 is dedicated to the numerical simulations of complete chaos synchronization between the Ikeda models. The discussions and conclusions of the results are presented in Section 4.

## II. Star network topology and complete synchronization

In this paper nodes are described by the Ikeda delay differential equations [19-20] A central node (hub) is connected to the peripheral nodes in a star-like manner [21] (spoke-hub configuration) Lines connecting the nodes are called links. Fig. 1 shows the case of the star networks topology. The central hub plays a role of conduit of message transmission.

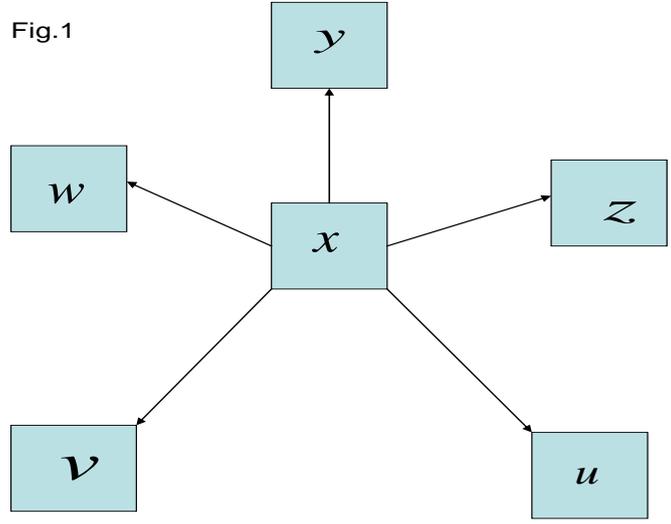

Fig.1 Schematic description of the case of the star networks topology.

Ikeda model [19] was introduced to describe the dynamics of an optical bi-stable resonator, playing an important role in electronics and physiological studies and is well-known for delay-induced chaotic behavior, see e.g. [19] and references therein. Later it was established that the Ikeda model or its modifications can be used to describe the dynamics of an opto-electronical, an acousto-optical systems and even the dynamics of the wavelength of lasers [20]. In overall Ikeda model is adequate for describing:
1) The propagation of laser intensity in a optical resonator with x being a phase shift in a nonlinear medium in the resonator; 2) Wavelength chaos; 3) For some solid state lasers, low pressure $CO_2$ lasers (the so-called B class lasers [22-23]; 4) Opto-electronic systems, where $x$ is the voltage fed to the KDP modulator; 5) Autoimmune diseases; 6) Neuronal models, etc.

Below we consider the case of the star network topology for the chaotic Ikeda model:

$$\frac{dx}{dt} = -\alpha x + m_1 \sin x(t-\tau), \qquad (1)$$

$$\frac{dy}{dt} = -\alpha y + m_2 \sin y(t-\tau) + m_3 \sin x(t-\tau), \qquad (2)$$

$$\frac{dz}{dt} = -\alpha z + m_4 \sin z(t-\tau) + m_5 \sin z(t-\tau), \qquad (3)$$

$$\frac{du}{dt} = -\alpha u + m_6 \sin u(t-\tau) + m_7 \sin x(t-\tau), \qquad (4)$$

$$\frac{dv}{dt} = -\alpha v + m_8 \sin v(t-\tau) + m_9 \sin x(t-\tau), \qquad (5)$$

$$\frac{dw}{dt} = -\alpha w + m_{10} \sin w(t-\tau) + m_{11} \sin x(t-\tau). \qquad (6)$$

In the original Ikeda model [19] $x$ is the phase shift experienced by the electric field in the nonlinear medium in the optical resonator. $\tau$ is the feedback time-delay. $m_1$ -is the feedback strength.

It should be noted that in the opto-electronic, electro-optical and acousto-optical systems dynamical variables are proportional to the voltage fed to a modulator [20]; $\tau$ is the feedback loop time delay (the propagation time of the light in the ring cavity). We will take the feedback time delays and connection delay times the same. $m_1, m_2, m_4, m_6, m_8, m_{10}$ are proportional to the intensities of the light injected into the optical resonator( feedback strengths. $m_3, m_5, m_7, m_9, m_{11}$ are the coupling strengths between the nodes. We also consider the case of identical relaxation coefficients.

By studying synchronization errors between the state variables $x, y, z, u, v, w$ one can find the existence conditions for the complete synchronization

$$x = y = z = u = v = w \qquad (7)$$

are:

$$m_1 = m_2 + m_3, m_2 = m_4 = m_6 = m_8 = m_{10}, m_3 = m_5 = m_7 = m_9 = m_{11} \qquad (8)$$

It is also straightforward to derive the stability condition, see e.g. [18] and references therein:

$$\alpha > |m_2| \qquad (9)$$

### III. Numerical simulations

This section numerically demonstrates that how the analytical findings of the previous Section are validated. We simulate the numerical calculations with the help of MATLAB 2008b. Synchronization quality is characterized by the correlation coefficient C between the dynamical variables. $C = 1$ means perfect complete synchronization.

We simulate Eqs. (1-6) for the following set of parameters

$$\alpha = 1.1, \tau = 6, m_1 = 5, m_2 = m_4 = m_6 = m_8 = m_{10} = 0.5,$$

$$m_3 = m_5 = m_7 = m_9 = m_{11} = 4.5$$

The initial states used in this work are:

$$x(0) = 2, y(0) = 2.1, z(0) = 2.2, u(0) = 2.3, v(0) = 2.4, w(0) = 2.5.$$

Fig. 2 depicts dynamics of the state variable $x$.

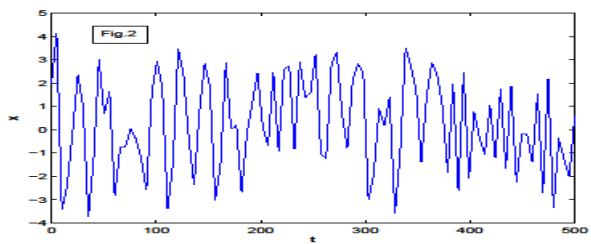

Fig. 2 Dynamics of variable $x$ in time for the set of parameters presented in Section 3. Dimensionless units.

Fig. 3 demonstrates error dynamics $x - w$ with time. It is seen that after short transients error $x - w$ approaches zero.

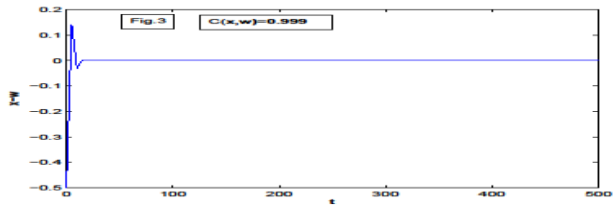

Fig.3 Dependence of $x - w$ on time t for the set of parameters presented in Section 3. $C = 0.999$ is the correlation coefficient between $x$ and $w$. Dimensionless units.

Fig. 4 presents the dependence between variables $y$ and $w$. This dependence is in full agreement with the theory and numerical simulations: the dependence is linear: $y = w$

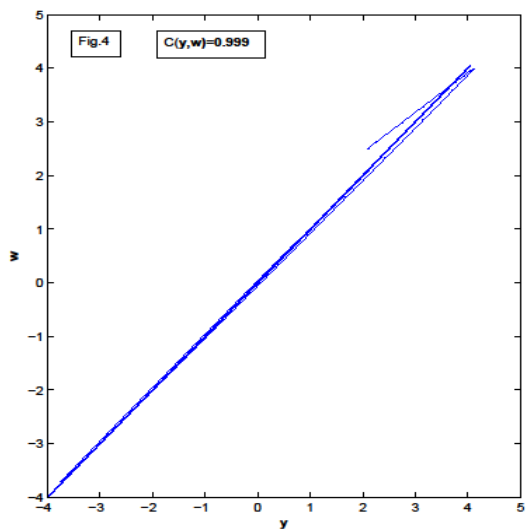

Fig. 4 Dynamics of variable $y$ versus dynamics of variable $w$ for parameters as in Section 3. The correlation coefficient $C$ between $y$ and $w$ is $C = 0.999$. Dimensionless units.

Fig. 5 shows the time dependence of error $z - w$. It is evident that with time $z$ approaches $w$.

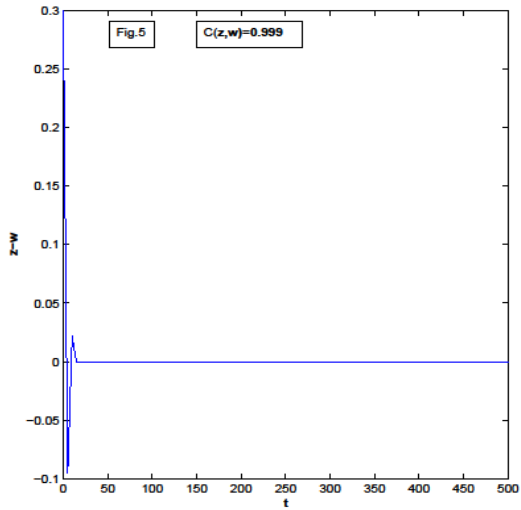

Fig.5 Dependence of $z - w$ on time t for parameters as in Section 3.
The correlation coefficient $C$ between $z$ and $w$ is $C = 0.999$. Dimensionless units.

Finally we numerically calculate the correlation coefficients between all the possible combinations of the state variables $x, y, z, u, v, w$. These results are presented in the Table 1.

Table 1: Correlation coefficients $C$ between the central node and peripheral nodes $x, y, z, u, v, w$ with all the possible combinations.

| $C(x,y)=1$ | $C(x,z)=0.999$ | $C(x,u)=0.999$ |
| --- | --- | --- |
| $C(x,v)=0.999$ | $C(x,w)=0.999$ | $C(y,z)=1$ |
| $C(y,u)=0.999$ | $C(y,v)=0.999$ | $C(y,w)=0.999$ |
| $C(z,u)=1$ | $C(z,v)=0.999$ | $C(z,w)=0.999$ |
| $C(u,v)=1$ | $C(u,w)=0.999$ | $C(v,w)=1$ |

IV. Discussions and Conclusions

In this work we study the case of a star networks topology on the example of Ikeda model –prominent multidisciplinary system. The existence and stability conditions for the complete synchronization between the constituent systems are derived. Numerical simulations fully support the theory. Possible security implications for the computer networks are emphasized. We have shown that high level of synchronization occurs between all the possible combinations of the constituent systems in spite of facts there are no direct connections between some Ikeda systems-peripheral nodes.

In this paper the bridge between chaos control method-synchronization and widely used in the computer networks of the star topology is built. This is a original approach to connect chaos synchronization and computer networks topology. As mentioned earlier synchronization is vital in chaos-based communication system to decode the transmitted message [1]. Chaos based communication approach could provide an extra layer of security in data exchange between communicating computers.

We also briefly dwell on the advantages and disadvantages of the star networks topology architecture [21]. Studied in this paper the star topology is used in the local area networks, such as home networks, Automated Teller Machine networks in bank transactions, hospital networks, Closed-Circuit TV networks mainly for surveillance and security purposes. Among advantages of star topology one can emphasize [21]:1) if one peripheral node or its connection fails the other nodes are still operable; 2) new devices can be removed and added without disturbing the entire network; 3)applicable for large network; 4 ) easy to manage, etc. Among the drawbacks of the star topology one should mention [21]: 1) central hub is the bottleneck; 2) very expensive, etc. Wiring the star networks topology can be implemented with the optical fiber cable and twisted pair cable. In the former case high speed communication can be achieved. Usually transmission rates for star topology are less than 1Gbit/s [24].

In Terahertz communications [25] an effective data exchange rates exceeding 1Tbit/s can be realized. In the case of optical cables between the networks data rates could be even higher than 1Tbit/s. This is due to the fact that wave division multiplexing approach [26] can be exploited. In this approach data packets can be sent using multiple wavelengths. The studied in this paper configuration could serve as a building block for much more complex networks and computer architecture.

Figure captions

Fig.1 Schematic description of the case of the star networks topology.

Fig. 2 Dynamics of variable $x$ in time for the set of parameters presented in Section 3. Dimensionless units.

Fig.3 Dependence of $x-w$ on time t for the set of parameters presented in Section 3/

$C = 0.999$ is the correlation coefficient between $x$ and $w$. Dimensionless units.

Fig. 4 Dynamics of variable $y$ versus dynamics of variable $w$ for parameters as in Section 3. The correlation coefficient $C$ between $y$ and $w$ is $C = 0.999$. Dimensionless units

Fig. 5 Dependence of $z - w$ on time t for parameters as in Section 3.
The correlation coefficient $C$ between $z$ and $w$ is $C = 0.999$. Dimensionless units.

## References

[1] E. Schoell and H.G. Schuster, *Handbook of Chaos Control*, 2-nd edition, Wiley-VCH (2008).

[2] A.T. Azar and S. Vaidyanathan, *Chaos modeling and control systems design*, Springer (2015).